# Agent-based opinion formation modeling in social network: a perspective of social psychology

Xicheng Yin[a], Hongwei Wang[a,*], Pei Yin[b], Hengmin Zhu[c]

[a] School of Economics and Management, Tongji University, Shanghai 200092, China
[b] Business School, University of Shanghai for Science and Technology, Shanghai 200093, China
[c] School of Management, Nanjing University of Posts and Telecommunications, Nanjing 210003, China

**Abstract**

Most previous works on opinion modeling lack the simultaneous study of individual mental activity and group behavior. This paper is motivated to propose an agent-based online opinion formation model based on attitude change theory, group behavior theory and evolutionary game theory in the perspective of sociology and psychology. In this model, there are three factors influencing the persuasion process, including credibility of the leaders, characteristic of the recipient, and group environment. The proposed model is applied to Twitter to analyze the influence of topic type, parameter changing, and opinion leaders on opinion formation. Experimental results show that the opinion evolution of controversial topic shows greater uncertainty and sustainability. The ratio of benefit to cost has a significant impact on opinion formation and a moderate ratio will result in the longest relaxation time or most unified global opinions. Furthermore, celebrities with a large number of followers are more capable of influencing public opinion than experts. This paper enriches the researches on opinion formation modeling, and the results provide managerial insights for business on public relations and market prediction.

---
* Corresponding Author.
  E-mail address: hwwang@tongji.edu.cn (H. Wang), xichengyin@tongji.edu.cn (X. Yin).

**Keywords**

Opinion formation, social psychology, attitude change, game theory, social network

# 1. Introduction

Extant studies show that opinion formation in social network is of great importance in various fields such as word of mouth marketing [1,2], political election, [3,4] and social governance [5-7]. Since the study of online opinion formation involves both psychological behavior and network dynamics, it has been a hot topic in sociology and nonlinear physics [8]. In general, one's opinion represents his attitude or standpoint towards a certain object, and opinion dynamics aims to reveal how social opinions evolve and converge by defining different interaction mechanisms from individual levels [9,10]. Although there have been a large amount of literatures on opinion model, most of them focus on nonlinear physics or statistical physics methods [8], which are incapable of illustrating individual mental activity and group behavior in details. In fact, the evolution of group opinion is a complex and holistic process, and the thoughts and behavior of each individual is necessary to be considered. Therefore, it is of great significance to model group's opinion formation at the individual level.

To tackle the problem, prior works adopt agent-based modeling, which is a relatively successful method used in social dynamics and many opinion models are based on it [8]. Since French put forward an opinion formation model as early as 1956 [11], a number of agent-based models have been proposed, which could fall into one of the two categories, continuous [12,13] or discrete [14-17], according to the opinion variables are defined. For either continuous or discrete, the general framework of opinion models could be summarized as follows: the agents' opinions in a given network are initialized; at each time step, agents interact with each other under certain interaction rule and then decide whether to change their opinions; finally all agents' opinions tend to reach an equilibrium state [10].

Regarded as a social man, rather than a node, the agent's behavior is affected not only by interactions with others but also by group environment [18,19]. Sociology and psychology theories are also important theory evidence to describe microcosmic individual interaction and macroscopic group behavior in the process of opinion formation. Specifically, theories of behavioral psychology like stimulus-response theory [20] (later improved as stimulus-object-response [21]) may be employed to explain individual behaviors in opinion formation: if we regard opinions received as a stimulus, individual response is to decide whether to change his opinion. And the famous attitude change model proposed by Hovland is a theory basis to simulate individual mental state: stick to the point of view or transform the attitude [22,23]. Moreover, human behavior including mental activity is the result of the individual interaction, and is also influenced by group environment, which has been confirmed by the sociology researches, *e.g.* Asch experiment [18] and Lewin's Field Theory [19].

As a theoretical method of group behaviors, evolutionary game theory has been widely applied to group dynamics [8,24]. Evolutionary games focus on how individuals continuously achieve maximum returns in the process of repeated games. This idea of dynamic evolution could be applied in opinion formation if we consider agents with different opinions as players taking different strategies [25]. In evolutionary games, the behavior of each agent along with its interaction with the group are described separately, and a transition from individual behavior to group behavior would take place when agents constantly adjust their strategies, *i.e.* opinion.

This paper proposes an online opinion formation model based on attitude change theory, group behavior theory, and evolutionary game theory. Whether a user follow his leader's opinion depends on three aspects: credibility of leaders, characteristic of recipient, and the situation. To be specific, credibility of leaders consists of two factors, *i.e.* expertise and trustworthiness, whereas characteristic of recipient can be measured by his original attitude and stubbornness. Affected by surrounding situation, each user will dynamically update their binary opinion strategy to adapt to the group because of our assumption that both of the two connected agents receive a benefit if they have the same opinion, or otherwise they both pay a cost [25].

Based on Twitter's data, we analyze the influence of topic type, parameter changing, and opinion leaders on the opinion formation. Results show that the opinion evolution of controversial topic shows greater uncertainty and sustainability. The ratio of benefit to cost has a significant impact on the opinion formation and a moderate ratio will result in the longest relaxation time or most unified global opinions. Besides that, celebrities are more capable of influencing public opinion than experts.

The remainder of this paper is organized as follows. Section 2 summarizes the theoretical background, and section 3 proposes the opinion formation model. The simulation results are provided in Section 4, followed by discussion in Section 5, and concluding remarks in Section 6.

## 2. Theoretical background

*2.1. Opinion models*

Opinion model aims to gain a fundamental understanding of how social opinions evolve and converge by defining different interaction mechanisms from individual levels [9,10]. Nowadays, Agent-based modeling is a commonly used method in social dynamics, thus being applicable by most opinion models [8]. French proposes an opinion formation model as early as 1956 [11], where continuous opinion is measured with a ratio scale and the effect of network connectedness on opinion changes in the group is illustrated as well. Generally, agent-based models could fall into one of the two categories, continuous or discrete, according to the opinion variables are defined. Deffuant model [12] and HK model [13] are typical continuous opinion models, and discrete opinion models include Sznajd model [14,15], Voter model [16], and Galam model [17]. These early opinion models regard all individuals as the same, so later studies attempt to improve these models by considering the heterogeneity of agents, *e.g.* stubbornness [26,27] and social power [28-30].

Among different models, the rules of opinion interaction between agents are not the same. Voter model assumes that an individual chooses a neighbor's opinion randomly as his or her own at each time step [16]. Galam model assumes that the agent tends to select the most common opinion in a group [17]. As continuous opinion models, Deffuant model [12] and HK model [13] are based on bounded confidence strategy, which holds that agents could communicate only if their opinion difference is less than a threshold.

Although some models consider the social attributes of agents or design the opinion interaction rules according to real social scenes, however, most of them put emphasis on modeling and simulation with nonlinear physics or statistical physical methods. Considering the opinion formation involves complicated individual mental status and group environment, it is necessary to model it systematically from the perspective of sociology and psychology.

*2.2. Theories of sociology and psychology*

Behavioral psychologist proposes the stimulus-response theory to describe complicated behaviors of human being [20], holding that human behavior in general is a response to a stimulus and there is a regular relationship between environmental stimuli and behavioral responses. Considering the intermediate factor (*i.e.* internal change caused by external stimuli) influencing individual's response, neo behaviorism introduce a mediating variable (object) into the stimulus-response scheme, which helps explain the individual behavior difference [21]. Different from behaviorism neglecting the role of human beings (treat a person as a passive object connecting stimulation and response), humanistic psychology represented by Maslow tends to regard people as the primary and core elements of communication [31]. As for the persuasion process, since changing one's mind is complicated involving several factors, *e.g.* opinion source, self-confidence and communication situation. To illustrate the transformation of attitude, Hovland puts forward the famous attitude change model, where persuader, the person who is being persuaded, persuasive messages, and situation are four essential elements associated with attitude change [22,23].

Sociology researches, *e.g.* Asch experiment [18] and Lewin's Field Theory [19], assert that human behavior results from individual interaction, and is also influenced by group environment. Symbolic interactionism, the beginning of the study of group communication, stresses that human beings need to cooperate with each other in the group whether for survival or psychological needs [32]. Asch conformity experiment explores if and how individuals yielded to or defied a majority group and such influences on beliefs and opinions. The results indicate that some people are willing to follow the group opinion even when it is inconsistent with their feelings [33-35]. In Lewin's Field Theory [19], each individual can be regarded as a field, and behavior is derived from the "psychological field", which is formed by the people and environment.

Consequently, considering the involvement of opinion disseminators and recipients in a whole system, microcosmic individual interaction and macroscopic group behavior both need to be studied in opinion formation model, and theories of sociology and psychology provide theoretical basis.

*2.3. Evolutionary game theory*

Game theory is widely used in the group behavior studies [24]. Different from the full-rationality based game theory, evolutionary game theory holds that the equilibrium state in the group can be achieved by individuals' constant attempt of adapting to the environment. Thus, evolutionary game focuses more on the dynamic change of agents' strategies.

Strategic game approach was first introduced to opinion model by Di Mare and Latora [36], and many scholars have attempted to combine evolutionary game theory with opinion model ever since [10,25,37]. In Di Mare and Latora's model, a player would gain a positive payoff if he convinces others to change their opinion; otherwise a negative payoff. Based on social conformity theory [38], Ref. [25] assumes that two connected agents will both receive a benefit if they hold the same opinion; otherwise the both will pay a cost.

In most evolutionary game models on opinion formation, a player often adjusts his strategy by calculating neighbor's payoff. For example, Ref. [37] assumes that every time an agent calculates the probability distributions of the opponent's reactions based on the history data, and then selects a best strategy with the highest utility. In Ref. [25], the payoffs of an agent $x$'s neighbors are computed and the opinion of whom with the highest payoff is most likely to be imitated by the agent. However, it may not be the case in real interactive context because information asymmetry makes it impossible for individual to obtain or calculate other's payoff. In online social networks, a user's opinion is usually affected by the "psychological field", which is composed of his own thoughts along with neighbors' opinions.

## 3. The proposed model

In directed social network such as Twitter or Instagram, the following relation is denoted by a directed link from follower to leader. In such network, the opinion spreads from a leader to his followers. We consider the opinion as binary, which represent two opposite attitudes towards particular topic, either yes or no. Furthermore, the model is also applicable to undirected network by changing the opinion receiving mechanism from unidirectional to bidirectional.

The proposed model consists of two parts: individual persuasion and group evolution. The former aims to illustrate the persuasion process of user-user pairs in details, whereas the latter focuses on the dynamic evolution of group behaviors, *i.e.* opinion strategies. One hypothesis of the model is that the user's opinion could be received without delay by his every follower (or friend) in social network.

*3.1. Individual persuasion*

Enlightened by stimulus-object-response theory [20], if we regard the received opinion as a stimulus, then individual's response is to decide whether to transform his opinion. Once receiving an opinion from leaders, individual will analyze such information with his own thoughts to obtain a more comprehensive cognition than before (Fig. 1). This cognition will determine whether to strengthen or weaken his original opinion, thus affecting his subsequent decision: stick to original opinion or change the attitude.

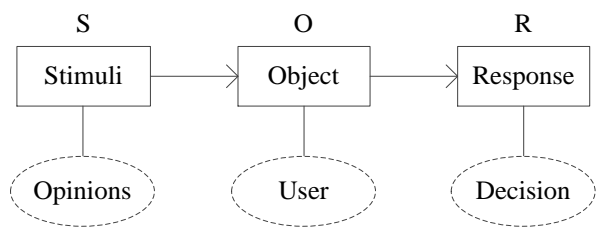

**Fig. 1.** Stimulus-object-response process of opinion formation

Based on Hovland's attitude change theory [22,23], we assume that whether a user is persuaded by his leaders depends on three aspects (Fig. 2): credibility of leaders, characteristic of recipient (person who received opinions), and the situation (group environment). The first two factors play a part in individual persuasion of user-user pairs, whereas the factor of situation works in group evolution. In addition, due to the similar content (the object of discussion) and diffusion behavior (post or repost) of the same topic in social network, the persuasive messages in Hovland's theory is not considered in this model.

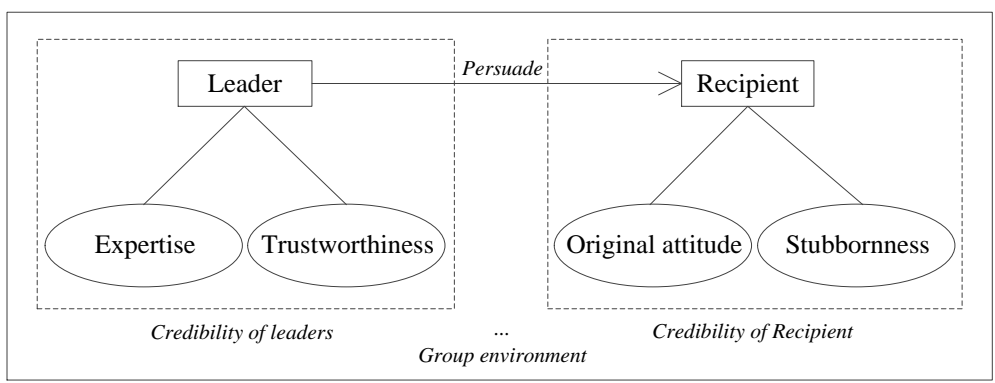

**Fig. 2.** Opinion persuasion model based on Hovland attitude change theory

### 3.1.1. Credibility of leaders

In Hovland's study of how one's attitude is affected by others, characteristic of the interacting object is regarded as an important variable, mainly referring to credibility and attractiveness. In opinion formation, agent's credibility is a crucial factor to maximize the influence power [39]. Also, since the attractiveness of an online user is not the same to each follower and thus is hard to define, we only adopt the credibility as the characteristic variable. The higher the credibility of opinion source is, the more likely the recipient is persuaded. According to attitude change theory, the credibility of a leader depends mainly on his expertise and trustworthiness.

(a) Expertise

Expertise means extensive knowledge, ability or experience in a particular field [40]. Generally, people with high expertise of a certain field are more likely to convince others. Regarding the influence of nuclear power stations, for example, users are more inclined to accept a scientist's opinion than a singer's. This assumption is supported by Aronson's laboratory experiment [41], in which communication attributed to a highly credible source shows greater effect on changing readers' opinions.

Due to the limited number of experts and the moderate level for the crowd's expertise, we assume that the expertise obeys normal distribution in social network. Normal distribution is widely used to represent real-valued random variables [42]. It describes the characteristics of many social, economic and natural phenomena, such as examination result, IQ scores, height of a population. According to central limit theorem, physical quantities that are expected to be the sum of many independent processes often have distributions that are nearly normal [43], and people's expertise in a certain field just meet this condition.

Let the expertise of user $i$ be described by $e(i)$. And all users' expertise obeys normal distribution $e(i) \sim N(\mu, \sigma^2)$, where $\mu$ denotes the intermediate level of overall expertise, and $\sigma^2$ measures the degree of dispersion. In later simulation, the distribution of expertise is re-simulated for different topics.

(b) Trustworthiness

Trustworthiness is the state or quality of being trustworthy or reliable. In online social networks, the most intuitive reflection of a user's trustworthiness is his social impact on others. In this model, trustworthiness indicates a person's reliability at any time, whereas expertise just represents the authority of a person on a particular topic. For instance, a famous singer has an eminent status and thus deserves others' trust, but he is not as high as a scientist in term of expertise when talking about nuclear issues.

We use in-degree (or degree in undirected network) to describe a node's trustworthiness and the reasons vary. Ref. [29] finds that the degree of an agent is better to measure social power in scale-free networks. In networking science, degree centrality is an efficient and simple property to measure the importance of a node because nodes with high degree can influence more nodes effectively due to having the greatest number of neighbors [44,45]. Moreover, in social networking sites, whether a user is trustworthy is largely determined by the number of his followers because the followers are online visible to everyone, and a famous user usually is popular, attracting as many as 10 million fan-followers. Thus, degree centrality is a reasonable property to measure trustworthiness of an online user.

Denoted by $t(i)$ the trustworthiness of user $i$, we have

$$t(i) = k_i^{\partial} \tag{1}$$

where $k_i$ denotes the in-degree of node $i$, and $\partial$ is coefficient.

*3.1.2. Characteristic of recipient*

The persuasiveness of opinion depends largely on the credibility of leaders, as well as the characteristic of opinion recipient. From the recipient's perspective, his original attitude and stubbornness will affect the persuasiveness.

(a) Original attitude

In psychological view, an attitude is an acquired or predisposed mental state regarding an object with some degree of negativity, which is perceived from a social or personal stimulus [46]. Generally, a person's attitude is stable and hard to change easily [47]. When a user receives an opinion, he will compare it with his original opinion and decide whether to update his strategy according to the payoff of current strategy.

(b) Stubbornness

Recall that agents in social network differ in many ways, one of which is stubbornness. Normally, the more stubborn a user is, the harder he is to accept others' opinions. In this model, we have an assumption that stubbornness is proportional to expertise. Special skills or knowledge can build up a person's confidence in a particular subject, so he is hardly convinced of others' opinions. Petty *et al.* find that people's assessment of the validity of their own thoughts has an impact on persuasion, and confident thoughts are more easily to guide attitudes. In their experiment, if a person is confident enough in his ability, he will be less likely to change his mind [48]. Besides that, since stubbornness is related to expertise, a person's stubbornness may vary regarding different topic types. For example, a scientist tends to stick to his original opinion on scientific issues, whereas a singer shows more confidence when it comes to music.

Thus, the stubbornness $s(i)$ of user $i$ is described by

$$s(i) = \beta \cdot e(i) \tag{2}$$

where $e(i)$ represents expertise and $\beta$ denotes the coefficient.

### *3.2. Group evolution*

Opinion adoption depends largely on the individual interaction between pairs of users, and is affected by the group as well. Group evolution focuses on the influence of all leaders on a user and the dynamic evolution of group behaviors, *i.e.* opinion strategies.

### *3.2.1. Situation*

Hovland's theory believes that the persuading process takes place not only between persuader and recipient, but also under certain situation, which mainly refers to diverse information, distracting environment and repeated information. Regarding the online social activity, we can hardly capture the distracting environment because the interaction actually occurs online rather than face-to-face. Instead, diverse and repeated information can be available as an influence of one's leaders. Therefore, this model regards the situation of a persuasion as group environment.

The direct impact of the group environment to users is social conformity, which is a type of social influence that results in a change of behavior or belief in order to fit in with a group. Conformity experiments indicate that some people are willing to follow the crowd even it is inconsistent with their own feelings [18,33-35]. Due to social conformity, people tend to adopt a dominant strategy under a specific circumstance. For instance, smokers obey no-smoking rule in public places. Similarly, when a user shares an opinion with his neighbors in a social network, he will get a sense of identity or a sense of belonging if he agrees with most other people. Otherwise, he will suffer peer pressure of disagreeing with others. In this paper, opinion interaction is considered as a game. As a result of social conformity theory, we assume that the connected agents in a user-user pair will both receive a benefit $b$ if they share the same opinion, or otherwise they will both pay a cost $c$ [25]. The payoff matrix for each user in an opinion interaction is given in Table 1.

**Table 1**
Payoffs for users with binary opinion.

|  | Opinion A | Opinion B |
|---|---|---|
| Opinion A | $b$ | $-c$ |
| Opinion B | $-c$ | $b$ |

When calculating the payoff of user $i$ at a certain time, we need to consider the effects of the characteristics of leaders and recipient. For recipient $i$, if his opinion is the same as one of his leader $x$, he will gain a benefit $b$, or otherwise pay a cost $c$. Besides that, the benefit or cost is affected by the characteristics of leaders and recipient as well, and thus need to be further adjusted.

(a) The effect of leader's expertise

As illustrated above, people with high expertise of a certain field are more likely to succeed in persuading others. Given that recipient $i$ and leader $x$ have the same opinion, recipient $i$ will feel more sense of identity if leader $x$ possesses higher expertise. As a result, an additional benefit $\omega_1 \cdot e(x) \cdot b$ is introduced, where $e(x)$ denotes the expertise of leader $x$. The benefit rises in proportion to $e(x)$ and $\omega_1$ is the coefficient that measures the effect of expertise on benefit (or cost). Similarly, if recipient $i$ and leader $x$ have different opinions, recipient $i$ will suffer heavier peer pressure if leader $x$ possesses higher expertise, which leads to an additional cost $\omega_1 \cdot e(x) \cdot c$.

(b) The effect of leader's trustworthiness

Similar to expertise, leader's trustworthiness also affects recipient's benefit, and higher trustworthiness indicates greater social influence on others. $\omega_2 \cdot t(x) \cdot b$ denotes the additional benefit brought by the trustworthiness of leader $x$, whereas the extra cost resulted from the pressure of disagreeing with a trustworthy leader is $\omega_2 \cdot t(x) \cdot c$, where $t(x)$ is the trustworthiness of leader $x$, and $\omega_2$ measures the effect of trustworthiness on benefit (or cost).

(c) The effect of recipient's stubbornness

Recall that the more stubborn a user is, the more hardly he will accept others' views. For a stubborn man, his perceived pressure caused by disagreement with others may be less, compared with someone with lower stubbornness, for he does not care much about others' feeling. As a consequence, stubbornness helps reduce the cost to a certain extent, meaning that stubbornness has a negative impact on cost. And we thus formulate $-\omega_3 \cdot s(i) \cdot c$ to measures the effect of stubbornness on cost.

Finally, the payoff matrix for each user in an opinion interaction is adjusted (see Table. 2).

**Table 2**
Adjusted payoffs for users with binary opinion.

|  | Opinion A | Opinion B |
|---|---|---|
| Opinion A | $b + \omega_1 e(x)b + \omega_2 t(x)b$ | $-[c + \omega_1 e(x)c + \omega_2 t(x)c - \omega_3 s(i)c]$ |
| Opinion B | $-[c + \omega_1 e(x)c + \omega_2 t(x)c - \omega_3 s(i)c]$ | $b + \omega_1 e(x)b + \omega_2 t(x)b$ |

Now suppose that there are two types of opinions: opinion $A$ and opinion $B$, denoted as +1 and -1 respectively. $O_i$ and $O_x$ are the opinion value ($\pm 1$) of user $i$ and $x$ respectively, and their product is -1 if they have different opinions. $p(i)$, the payoff of recipient $i$ at a particular time, is the sum of payoff obtained by receiving opinion from each leader. We formulate $p(i)$ as follows.

$$\begin{cases} p(i) = \sum_{x \in \Omega_i^+} O_i O_x b' + \sum_{x \in \Omega_i^-} O_i O_x c' \\ b' = b + \omega_1 e(x)b + \omega_2 t(x)b \\ c' = c + \omega_1 e(x)c + \omega_2 t(x)c - \omega_3 s(i)c \\ s(i) = \beta e(i) \end{cases} \quad (3)$$

After deriving, we have Equation (4), where $\Omega_i^+$ represents the leader set having the same opinion as recipient $i$ and $\Omega_i^-$ denotes the leader set having different opinion.

$$p(i) = b \sum_{x \in \Omega_i^+} [1 + w_1 t(x) + w_2 e(x)] - c \sum_{x \in \Omega_i^-} [1 + w_1 t(x) + w_2 e(x) - w_3 \beta e(i)] \quad (4)$$

*3.2.2. Evolution*

In evolutionary game theory, agents are not completely rational, and the same is true for online users. In the process of opinion formation, the online users keep adjusting their cognition to adapt to public opinion. They will decide whether to change their strategy (opinion) at each time step. In general, a person's attitude is hardly to change because of attitude's stability, but if the current strategy (opinion) brings him more cost than benefit, he is willing to change.

Specifically, it is assumed that user $i$ holds opinion $A$ at time $t$, with a payoff of $p(i)$. When $p(i) < 0$, which means the benefit of current strategy is less than cost, user $i$ will change his strategy to $B$ at next time for more payoff, otherwise remain unchanged. Moreover, we need to note that a user's choice of strategy is made by calculating his own payoff instead of his leaders', which is more consistent with actual scene.

## 4. Simulation results

### 4.1. Data description

Our dataset comes from Twitter, provided by Social Computing Data Repository at Arizona State University [49]. There are about ten million nodes and eighty million follow relations. To refine the social network, we conduct a pruning process, by which leaf nodes are eliminated repeatedly until social networking features (average path length, average clustering coefficient, etc.) became obvious. The basic topological properties of the refined network are shown in Table 3, and its in-degree distribution approximately follows the power-law distribution (as shown in Fig. 3), which is a typical feature of online communities.

**Table 3**
Basic topological properties of the network.

| N | M | <k> | D | L | C |
|---|---|---|---|---|---|
| 12418 | 99087 | 7.979 | 15 | 4.747 | 0.132 |

Note: N and M are the total numbers of nodes and links, respectively; <k> denotes the average degree; D is the network diameter; L is the average path length and average clustering coefficient is denoted by C.

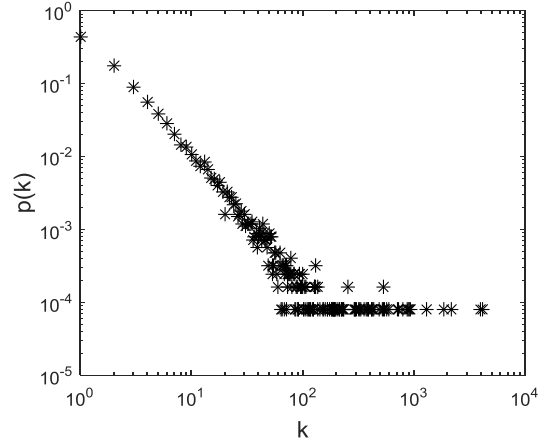

**Fig 3.** In-degree distribution of the experimental network in double logarithmic coordinate system. $k$ is the in-degree of nodes and $P(k)$ denotes the probability of a node with in-degree $k$.

*4.2. Simulation of opinion formation*

*4.2.1. Influence of topic type*

Online topics can fall into one of the two types, either controversial topic or one-sided topic. Controversial topic is usually the subject of intense public argument, disagreement, or disapproval, and thus arouses fierce discussions within a group, whereas one-sided topic, on the contrary, gets a unified opinion at the beginning.

Based on the proposed model, we simulate the opinion formation process in Twitter network above. To simplify, benefit $b$ and cost $c$ are both set as 1, and coefficient $\omega_1$, $\omega_2$, and $\omega_3$ are set the same because the impact of expertise, trustworthiness or stubbornness on benefit are treated equally. Coefficient $\omega_1$, $\omega_2$, $\omega_3$, $\partial$, and $\beta$ are initialized as 1, and all users' expertise obeys normal distribution, *i.e.* $e(i) \sim N(\mu, \sigma^2)$, with $\mu = 10$, $\sigma^2 = 0.25$. To adjust values measured on different scales to a notionally common scale, $t(x)$, $e(x)$, and $s(i)$ are normalized from 0 to 1.

Considering topic type, two topics are simulated to observe the different effects of topic type on opinion formation. The initial proportion of opinion $A$ of controversial topic is 55%, whereas that of one-sided topic is 75%. At the initial time $t = 1$, 55% (or 75%) of users are randomly selected to hold opinion $A$, with the others holding opinion $B$. We simulate opinion formation 100 times for each topic, and get the average results. Fig. 4 shows how the proportion of opinion $A$ and average payoff of the network vary over time.

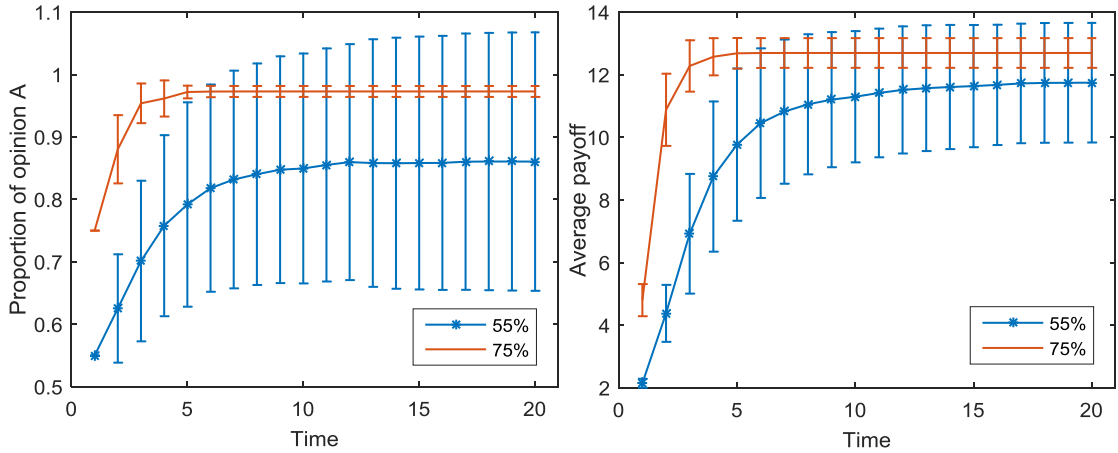

**Fig. 4.** The proportion of opinion $A$ (left) and average payoff (right) of all users varying with time step in Twitter network. Data shows the mean value with error bars corresponding to the standard deviation over 100 realizations.

Fig. 4 demonstrates that the proportion of opinion $A$ tends to increase rapidly and then becomes stable with low fluctuation. The average payoff of the network also keeps an upward tendency until stabilizing. In addition, the result deviation of controversial topic over 100 realizations is large, whereas that of one-sided topic is very small.

To examine the effect of topic type on opinion formation in detail, more cases of initial proportion of opinion $A$ is considered, with initial proportion ranging from 55% to 80% (Fig. 5). And relaxation time is used to describe the time when the proportion of opinion $A$ reaches a steady state. As illustrated in Fig. 5, the topic with higher initial proportion of opinion $A$ tends to reach the relaxation time earlier with a higher final proportion of opinion $A$.

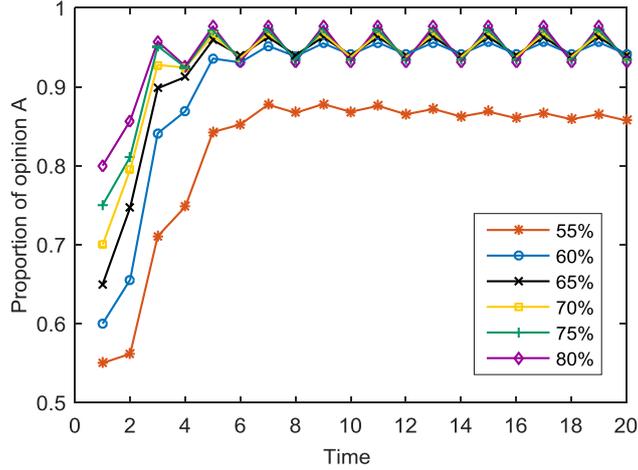

**Fig. 5.** The proportion of opinion $A$ varying with time step in Twitter network, with initial proportion ranging from 55% to 80%. Data shows the mean value over 100 realizations.

*4.2.2. Influence of parameter changing*

There are four parameters in the payoff formula: $b/c, \partial, \beta, \mu$, and $\sigma^2$. $b$ means the benefit when one's opinion is supported by one of leaders, and $c$ denotes the cost if he disagrees with one of leaders; $\partial$ is coefficient in Equation (1) for trustworthiness; $\beta$ denotes the coefficient in Equation (2) defining stubbornness; $\mu$ denotes the intermediate level of overall expertise, and $\sigma^2$ measures the degree of dispersion. Because of data normalization, $\mu$ and $\beta$ actually do not affect the results and thus they can be ignored. The effect of $b/c, \partial$, and $\sigma^2$ are analyzed with the initial proportion of opinion $A$ be 55%. The experimental result shows that only $b/c$ has a significant effect on the opinion formation, whereas the change of $\partial$ and $\sigma^2$ makes no difference. This is explicable because coefficient $\partial$, as an exponent, makes a user's trustworthiness grow with the increase of in-nodes (followers), but its value only affects the speed of this increasing relation. Coefficient $\sigma^2$ measures the dispersion degree of a group's expertise, and its value varies in different social communities. But it is obvious that a topic would finally make consensus in different communities, which is approved by previous experiments [10,25].

Basically, $b/c$ describes the relative change of benefit $b$ for cost $c$. For simplicity, the cost $c$

is constantly set as 1, and thus only the value of $b$ need to be under control. $b > 1$ means the benefit of sharing a same opinion is greater than the cost of holding different opinion. Fig. 6 shows how relaxation time and final proportion of opinion $A$ change with different values of $b/c$, and the result is the average of 100 times simulations.

Fig. 6 demonstrates that for given values of other parameters, there exists a particular value (around 1.6) of $b/c$ resulting in the longest relaxation time. Also, a moderate $b/c$ (around 1.0) maximizes the final proportion of opinion $A$ at relaxation time, in another words, gets the most unified global opinions within group.

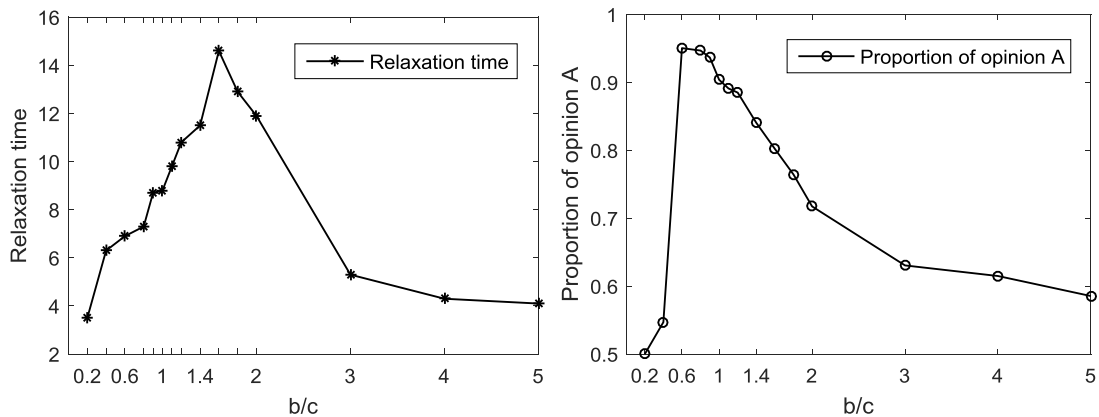

**Fig. 6.** The relaxation time (left) and final proportion of opinion $A$ (right) as a function of $b/c$ respectively, with the initial proportion of opinion $A$ be 55%.

*4.2.3. Influence of opinion leader*

How to guide the forming opinions to reach a consensus has received extensive attention from scholars and government [50], and opinion leader is an important way to guide the opinion. Generally, opinion leaders are influential members of a group or society to whom others turn for advice, opinions, and views [51,52]. The elites like entertainers, entrepreneurs, and politicians have more scope to shape and lead public opinion, and meanwhile the experts of a certain filed also lead public opinions on some topics [53]. Therefore, we identify opinion leaders according to in-degree and expertise, and examine the impact of opinion leaders on opinion formation by initially assigning opinion *A* to users with high in-degree or expertise and then comparing the result with previous ones.

Earlier in the paper, the initial 55% users with opinion *A* are selected randomly, whereas the 55% users chosen in this experiment contains top-500 users with the highest in-degree. We also conduct another simulation, where the top-500 users with the highest expertise are included in the initial user set. And these 500 opinion leaders account only for 4% of the group's population.

Fig. 7 shows the difference of whether or not the guidance of opinion leaders is conducted. It shows that the proportion of users holding opinion *A* increases with the guidance of opinion leaders, and the increment degree of guiding with high in-degree is much greater than that of guiding with high expertise. Besides that, the result illustrates that the standard deviation of opinion proportion reduces a little after the guidance with expertise, whereas the deviation decreases drastically when we guide the opinion with in-degree. Moreover, compared with expertise's guiding effect, the guidance with high in-degree accelerates the opinion consensus and gets much smaller standard deviation of opinion proportion at each time-step and finally the group opinions becomes more unified. Overall, the impact of opinion leaders on opinion formation is significant, and the effect of guidance with in-degree is much more obvious than that with expertise.

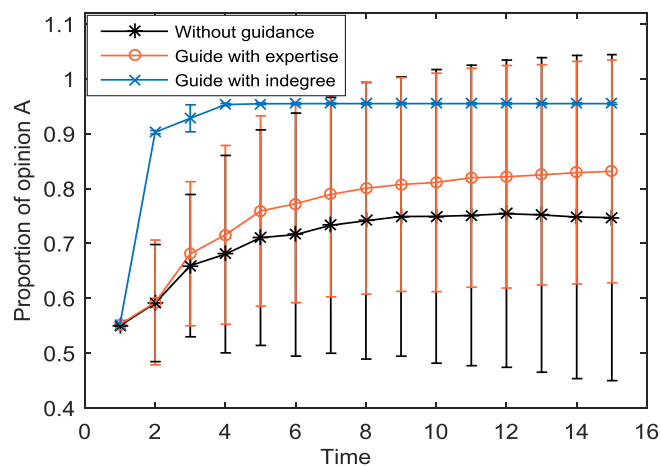

**Fig. 7.** The difference of whether the guidance of opinion leaders is conducted or not, with the initial proportion of opinion $A$ be 55%. Data shows the mean value with error bars corresponding to the standard deviation over 100 realizations

*4.3. Comparison*

We compare our model's results with state-of-the-art models put forward in recent years (Table. 4). Although models vary in agents' opinion adopting basis, the social group would finally reach a consensus on a certain opinion. In Ref. [25], there also exists an optimal ratio of cost to benefit, leading to the shortest consensus time, which further indicates that ratio of benefit to cost has a significant impact on opinion formation. And the consensus time decreases as the average degree of the network increases, which highlights the importance of in-degree in opinion guidance, as well. In addition, Ref. [10] and [54] show that the initial distribution of opinion is vital to the evolutionary result, which is agreed by our model. In the other three models, network properties, evolving rules, initial opinion distribution, and agents' rationality have influence on the opinion propagation, but these factors are inherent attributes, which is hard to be controlled as a guidance solution. In our model, users with high in-degree are proved to be effective to guide the opinion trend and it is easy for implementation. Actually, such guidance is being implemented by many business companies: sellers usually look for celebrities with a large number of fans to promote their products on Twitter. Overall, after taking sociology and psychology theories into account, the proposed model gives new supplement to opinion formation modeling, and some of the results are supported by previous studies.

**Table 4**
Results comparison with other opinion formation models.

| Model | Opinion updating basis | Result | | |
|---|---|---|---|---|
| | | Converge | Optimal $b/c$ | Effective guidance |
| This paper | Payoff | Yes | Exist | 1) High in-degree agents |
| Ref. [25] | Payoff | Yes | Exist | 1) Average in-degree |
| | | | | 2) Network size |
| | | | | 3) Less noise (irrational choices) |
| Ref. [10] | Payoff | Yes | Not discussed | 1) Evolving rule |
| | | | | 2) Network topology |
| | | | | 3) Initial opinion distribution |
| Ref. [54] | Bounded confidence | Yes | Not discussed | 1) Initial opinion distribution |
| | | | | 2) Opposite opinion |

## 5. Discussion

We simulate the opinion formation in three aspects: examining the effects of different topic type on opinion formation, analyzing the influence of parameter change, and considering the impact of opinion leaders on opinion formation as well. Each aspect provides evidence for better understanding of the spread of online opinions.

When the topic type is taken into account, we find that a group with more unified initial attitude leads to less time to reach consensus, which is consistent with the fact that people's initial opinions represent their first impression or attitude about the topic, and more unified initial attitude will contribute to reaching a consensus faster. This result has already been proved in the past experiments [10,53]. This is not always the case, however, for the reason that Fig. 4 shows a more obvious standard deviation and later relaxation time for controversial topics, indicating an uncertain and sustainable opinion formation process. Our result is also explicable in social media: a controversial topic often attracts two parties with opposite views to launch an intense discussion, and the two well-matched sides make the group's opinion evolution more unpredictable. Besides that, the deviation of topic spreading is also related to the distribution of users' initial opinions in the network. For example, the topic with more divergence is still possible to reach a more unified situation provided that some experts or any other important users happened to hold opinion *A* initially and continuously spreading this opinion to others. Moreover, the average payoff of the network also keeps an upward tendency until stabilizing, which means that the topic participants continuously acquire benefit (or sense of identity), and this might be a motivation for users to express their opinions online.

Through parameter adjustment, we find that the ratio of benefit to cost has a significant impact on the opinion formation and a moderate value of $b/c$ will result in the longest relaxation time or most unified global opinions. However, the exact value of benefit and cost in real social context remains unknown, since it varies from person to person. But generally the difference between benefit and cost is not large, as the satisfaction of being agreed by a leader and the peer pressure of disagreeing with a leader is similar if the leader's credibility and the receipt's characteristics are not considered. When $b/c$ is around 1, as illustrated in Fig. 5 and Fig. 6, the final proportion of opinion *A* at relaxation time keeps around 85% to 90% for controversial topic (with the initial proportion of opinion *A* is 55%). In other words, even for a topic with conflicting opinions, a group always reaches a consensus due to similar benefit and cost.

The simulation result also demonstrates the significant impact of opinion leaders on opinion formation, and the effect of guidance with in-degree is much more obvious than that with expertise. The underlying reason may be that users with high in-degree are capable of spreading out their tweets to more social groups through their followers' reposting, whereas some experts' valuable opinions are hard to broadcast with their limited followers. In Fig. 7, guidance with in-degree advances the relaxation time and makes the standard deviation of opinion proportion at each time-step very small. It means that the uncertainty and sustainability of the topic are weakened.

This result indicates that celebrities are of great help to guide public opinion in social network, and thus has become one of the most widely-used means of advertisement in industry.

## 6. Conclusion

This paper proposes an agent-based model to examine the process of online opinion formation in social network from the perspective of sociology and psychology. Regarding the agent as a social man, attitude change theory, group behavior theory, and evolutionary game theory are introduced to design the opinion interaction mechanism. To analyze both microcosmic individual interaction and macroscopic group behavior, leader's credibility, recipient's characteristic, and group environment are taken into account in the persuasion process of a user by his leaders. Besides, we apply evolutionary game theory to study dynamic change of users' opinions in topic propagation. Moreover, our model considers individual's heterogeneity by employing leader's expertise and trustworthiness as well as receipt's stubbornness.

Our experiments use Twitter's dataset to verify the proposed model. Although our model is targeted at directed network, it is also applicable to undirected network by simply changing the opinion receiving mechanism from unidirectional to bidirectional. We conduct different simulations to analyze the influence of topic type, parameter changing, and opinion leaders on the opinion formation process. The results show that more unified initial attitude in a group helps advance the time to reach consensus, and the process of opinion formation for controversial topic behaves with greater uncertainty and sustainability, often resulting in intensive argument. Among different parameters, only the benefit over cost ratio has a significant impact on the opinion formation process, and the given ration (around 1.6) leads to the longest relaxation time, whereas a moderate ratio (around 1.0) gets the most unified global opinions within group. The impact of opinion leaders on opinion formation is also significant, and the effect of guidance from celebrities is much more obvious than that from experts.

Opinion formation is a complex process involving individual mentality and group behavior. Different from previous works' emphasizing nonlinear physics or statistical physics methods, our method pays more attention to the combination of individual mental activity and group behavior, which is expected to provide insights for the study of opinion formation from the perspective of sociology and psychology. Further research will be conducted in the following aspects. A more sophisticated opinion model will be explored in order to analyze the hierarchy of opinion formation process. Opinion leaders will be identified through more complicated algorithms rather than simply considering in-degree and expertise. The proposed opinion model will be further applied to industrial practices, and help companies better understand how consumers' opinion spread on social media and adopt pertinent marketing measures in advance.